\documentclass[aps,twocolumn,prd,showpacs,nofootinbib]{revtex4}
\usepackage{amsmath}
\usepackage{graphicx}
\usepackage{dcolumn}
\usepackage{bm}
\usepackage{amssymb}
\usepackage{latexsym}

\def\setC{\mathbb{C}}

\def\setR{\mathbb{R}}

\newcommand{\ie}{\textsl{i.e.~}}

\newcommand{\GReCO}{${\cal G}\setR\varepsilon\setC{\cal O}$}

\def\spose#1{\hbox to 0pt{#1\hss}}
\def\lta{\mathrel{\spose{\lower 3pt\hbox{$\mathchar"218$}}
     \raise 2.0pt\hbox{$\mathchar"13C$}}}
\def\gta{\mathrel{\spose{\lower 3pt\hbox{$\mathchar"218$}}
     \raise 2.0pt\hbox{$\mathchar"13E$}}}

\begin{document}

\title{Reply to ``Bouncing Universes and their perturbations: remarks
on a toy model''.}

\author{J\'er\^ome Martin}
\email{jmartin@iap.fr}
\affiliation{Institut d'Astrophysique de
Paris, \GReCO, FRE 2435-CNRS, 98bis boulevard Arago, 75014 Paris,
France}

\author{Patrick Peter}
\email{peter@iap.fr}
\affiliation{Institut d'Astrophysique de
Paris, \GReCO, FRE 2435-CNRS, 98bis boulevard Arago, 75014 Paris,
France}

\date{June 14$^\mathrm{th}$, 2004}

\begin{abstract}
In this web note, we reply to a recent paper~\cite{nd1} confirming a
previous work of ours in which a cosmological bouncing phase was shown
to have the ability of modifying the spectrum of primordial
perturbations~\cite{jmpp}, but challenging its physical conditions of
validity.  Explicitly, Ref.~\cite{nd1}, besides pretending our Taylor
series expansion of the scale factor close to the bounce amounts to
choosing a family of polynomial scale factors, also claims that the
bounce affects the spectrum only if the mass scale of the scalar field
driving the dynamics is of the order of the Planck mass.  We show that
these objections are either misleading or incorrect since the minimum
size of the Universe $a_0$ (value of the scale factor at the bounce)
is either not physically specified, as required in a closed Universe,
or implicitly assumed to be the Planck mass. We calculate this mass
and obtain that, unsurprisingly, for a reasonable value of $a_0$, \ie
much larger than the Planck length, the scalar field mass is smaller
than the Planck mass.
\end{abstract}

\maketitle

In Ref.~\cite{nd1}, some comments and remarks were made concerning our
work of Ref.~\cite{jmpp}. The main result of Ref.~\cite{nd1} is an
independent confirmation of the fact that a bounce can affect
cosmological perturbations not only by an overall amplitude but also
by changing the spectrum, a result first obtained in Ref.~\cite{jmpp}.
However, Ref.~\cite{nd1} also criticizes the conditions of validity
under which the results of Ref.~\cite{jmpp} hold true, arguing that
the physical situations leading to such a spectral change are highly
specific and potentially problematic; the arguments produced are, we
believe, either misleading or erroneous, and in this short note we
want to correct them.

The purpose of our work~\cite{jmpp} was to discuss the propagation of
cosmological perturbations through a regular bounce, thus
concentrating explicitly in the immediate vicinity of the bounce
only. Since it is always possible to Taylor expand a regular function
on a finite interval, the choice of taking the scale factor as a
series in conformal time $\eta$ (a similar relation holding as a
function of cosmic time $t$)
\begin{equation}
\frac{a(\eta )}{a_0}\simeq 1+\frac{1}{2}\left(\frac{\eta }{\eta
_0}\right)^2 +\delta \left(\frac{\eta }{\eta
_0}\right)^3+\frac{5}{24}(1+\xi ) \left(\frac{\eta }{\eta _0}\right)^4
+\cdots 
\label{aseries}
\end{equation}
is the most general case that can be handled provided
$\eta\ll\eta_0$. This expansion was thus used in Ref.~\cite{jmpp} in
order to derive conclusions about the propagation of cosmological
perturbations through a bounce without specifying the model at all
times, in particular far from the bounce.

Apparently, Ref.~\cite{nd1} confused this Taylor series, valid only
close to the bounce, \ie for $\eta\ll\eta_0$, with an actual ``{\sl
family of polynomial scale factors}'', see Eq.~(2.3) of
Ref.~\cite{nd1}. This is made even clearer through the sentences:
``{\sl in order to relate the pre-bounce perturbation modes to the
post-bounce ones, one needs asymptotic regions where to define the
``in'' and ``out'' states, and hence a {\bf large $\eta$ behaviour}
... hence the assumption, {\bf made in [1]} in order to obtain an
asymptotically bounded $W(\eta)$ (see, e.g., equation (41) therein),
of the {\it polynomial} scale factor ...}'' (emphasis ours). Ref.~[1]
of this sentence means Ref.~\cite{jmpp} of the present note, whereas
$W(\eta)$ is the effective potential for scalar perturbations, denoted
$V_u(\eta)$ in Ref.~\cite{jmpp}. Let us comment these statements.

The ``in'' and ``out'' regions in the model under consideration exist
for $\eta\ll\eta_0$ provided $\eta_0$ is close to unity since in this
case, the width of the potential itself is of order $\sqrt{\Upsilon}
\equiv \sqrt{1-\eta_0^{-2}}\ll 1$, see Eqs.~(50) and (52) of
Ref.~\cite{jmpp}. Indeed, the effective potential rapidly goes to
$V_u(\eta\ll\eta_0)\to 3$ while the modes of cosmological interest are
such that $k^2 \gg 3$ (see discussion in Ref.~\cite{jmpp}). Therefore,
such ``in'' and ``out'' states can be defined even for
$\eta\ll\eta_0$: they happen to be just plane waves.  As a
consequence, the treatment of Ref.~\cite{jmpp} is valid whatever the
actual complete solution for the scale factor, again provided
$\Upsilon\ll 1$, and thus does not rest by any means on the assumption
the work of Ref.~\cite{jmpp} is credited by Ref.~\cite{nd1} (see also
FIG.~5 of Ref.~\cite{jmpp} which shows the explicit counter-example of
an effective potential calculated in an exact non-polynomial case,
$a(\eta )\propto \sqrt{1+\eta ^2/\eta _0^2}$, and compared to its
approximation for which $\xi =-2/5$). The embedding of our short
transition into a model that would be valid at all times was in fact
discussed around FIG.~10 in Ref.~\cite{jmpp}. Finally, let us note
that the purpose of using the fitting formula of Eqs.~(41) and (42) of
Ref.~\cite{jmpp} was in no way to obtain a bounded $V_u$ (or $W$) but
instead to get its actual value in the region where
Eq.~(\ref{aseries}) is valid. The actual approximation scheme for the
potential is in fact quite irrelevant (and in practice not used in
Ref.~\cite{jmpp}), what matters is the fact that the potential goes to
a constant (less than $k^2$) value in a short enough timescale, \ie
$\eta\ll\eta_0$ so that an asymptotic region is available. The crucial
point is that this property is valid not only for polynomial scale
factors, but for a much wider class (provided $\Upsilon$ is small
enough), as exemplified in FIG.~5 and discussed explicitly in the
caption. Moreover, one can find other non polynomial examples that
work in the same way~\cite{jmpp}, although it is probably possible to
cook up special cases for which the abovementioned property does not
hold.

Another point worth of a reply is the use of so-called ``rescaled
units'': if the spatial curvature $\mathcal{K}$ is normalized to unity
\{as done in Ref.~\cite{nd1}, see the ``1'' on the left-hand-side of
the Friedmann-Lema\^{\i}tre equation in the text above Eq.~(2.1) of
Ref.~\cite{nd1}\}, a perfectly consistent choice which is made indeed
very often, it is not possible to simply set the minimum value of the
scale factor $a_0$ as well to unity, as done in Eq.~(2.3) of
Ref.~\cite{nd1}. In fact, when the spatial curvature is non-vanishing,
as in the case at hand, the scale factor acquires a physical meaning
and thus cannot be rescaled at will when $\mathcal{K}=1$. Assuming it
is unity and that therefore all dimensionfull quantities are
implicitly expressed in units of $a_0$ is of course possible, but
seems to us to be a source of confusion and errors since obviously,
unless a value is assigned for $a_0$, no physical conclusion can be
drawn.

Let us illustrate our point by repeating the exercise consisting in
calculating the scalar field potential close to the bounce (we cannot
consider, given our framework, asymptotic expansions far from the
bounce, \ie $\eta \gg \eta _0$). Plugging a power series expansion
into another, it is not surprising that the result for the potential
$V(\varphi)$ is nothing but yet a power series in $\varphi-\varphi_0$,
where $\varphi_0$ is the field value at the bounce. Since the bounce
is assumed symmetric, an even function is necessarily obtained, \ie
\begin{equation} \label{Vphi}
V(\varphi) = V_0 + \frac{1}{2} M_\varphi^2
\left(\varphi-\varphi_0\right)^2 +\frac{1}{4!}  \lambda
\left(\varphi-\varphi_0\right)^4 +\cdots,
\end{equation}
with parameters $M_\varphi$ and $\lambda$ to be determined by the
coefficients appearing in the scale factor. Using Einstein and
Klein-Gordon equations for the background (\ref{aseries}), one
immediately gets that the square mass is negative, meaning the field
goes through a maximum of the potential as the bounce takes place,
and, to leading order in $\Upsilon$, that
\begin{equation}\label{m}
a_0 |M_\varphi|\sim \sqrt{\frac{-5\xi}{4\Upsilon}} \simeq 700,
\end{equation}
where in the last equality for the numerical application, we have
chosen, as in Ref.~\cite{nd1}, the values $\Upsilon = 10^{-6}$ and
$\xi = -2/5$. Clearly, as long as the value of $a_0$ has not been
provided, that is to say if one works with the units of
Ref.~\cite{nd1}, Eq.~(\ref{m}) is useless. Moreover, using the value
$a_0=1$, there is now the danger to conclude erroneously that the mass
parameter in the scalar field potential is larger than the Planck
mass. As a matter of fact, using natural units, it is clear that
provided the bounce occurs for a large enough scale factor, in
practice $a_0\gg 10^3 \ell_{_\mathrm{Pl}}$, the symmetry breaking mass
parameter in the potential can be much smaller than the Planck
scale. For instance, a natural choice could be the GUT scale ($a_0
\sim 10^5 \ell_{_\mathrm{Pl}}$), for which
\begin{eqnarray}
M_\varphi &\simeq &0.02 \times M_{_\mathrm{Pl}},
\\
\nonumber 
\end{eqnarray}
with $M_{_\mathrm{Pl}}=\ell^{-1}_{_\mathrm{Pl}}$ the Planck mass. In
connection with the discussion at the beginning of this note, let us
notice that this result is particularly interesting in view of the
fact that it is valid for a wide class of scale factors.

Finally, let us remark an additional mistake made in Ref.~\cite{nd1}
in which it is said that ``{\sl the total energy of the scalar field
... at the bounce is $\varrho(0)=3$ ... If one argues that what
matters is the value of the {\it total} energy of the field with
respect to the {\bf Planck energy}, then the restriction on $\Upsilon$
falls}'' (emphasis ours). In this particular case, and contrary to
what appears in Ref.~[7] of Ref.~\cite{nd1}, the units are now
explicitly Planckian. The correct statement should read
$\varrho(0)=3/a_0^2$ so that, again, if $a_0 \gg \ell_{_\mathrm{Pl}}$,
then $\varrho(0)$ can be much less than the Planck energy density. We
conclude that claiming the kinetic, potential or total energies of the
field must be Planckian near the bounce is incorrect.

\par

On a more general physical ground, it is clear that in a closed FLRW
universe for which the scale factor normalization is meaningful, there
is no hope to say anything relevant about the energies involved
without having specified the minimal size of the universe at the
bounce.

\vskip1cm

We wish to thank N.~Deruelle, N.~Pinto-Neto and J.-P.~Uzan, and
especially D.~J.~Schwarz for interesting remarks and careful reading
of the manuscript.

\end{document}